\pdfoutput=1
\documentclass[sigconf]{acmart}

\AtBeginDocument{%
  \providecommand\BibTeX{{%
    \normalfont B\kern-0.5em{\scshape i\kern-0.25em b}\kern-0.8em\TeX}}}

\setcopyright{acmcopyright}
\copyrightyear{2023} 
\acmYear{2023} 
\setcopyright{acmlicensed}\acmConference[ISCA '23]{Proceedings of the 50th Annual International Symposium on Computer Architecture}{June 17--21, 2023}{Orlando, FL, USA}
\acmBooktitle{Proceedings of the 50th Annual International Symposium on Computer Architecture (ISCA '23), June 17--21, 2023, Orlando, FL, USA}
\acmPrice{15.00}
\acmDOI{10.1145/3579371.3589049}
\acmISBN{979-8-4007-0095-8/23/06}

\usepackage{xspace}
\usepackage{graphicx}
\usepackage{caption}
\usepackage{multirow}
\usepackage{subcaption}

\newcommand{\ours}{\textsc{ArchGym}\xspace}
\newcommand{\oursnospace}{\textsc{ArchGym}\xspace}
\newcommand{\niparagraph}[1]{\vspace{1pt}\noindent\textbf{#1}}

\newcommand{\eg}[0]{\textit{e.g.}}
\newcommand{\Fig}[1]{Fig.~\ref{#1}}

\usepackage{enumitem} 

\newenvironment{itempacked}{%
\begin{itemize}[noitemsep,nolistsep,leftmargin=*]
}
{%
\end{itemize}
}

\newcommand{\rev}[1]{\textcolor{black}{#1}}
\usepackage{hyperref}
\begin{document}

\title{ArchGym: An Open-Source Gymnasium for Machine Learning Assisted Architecture Design}

\author{Srivatsan Krishnan}
\email{srivatsan@seas.harvard.edu}
\affiliation{%
  \institution{Harvard University}
  \city{Cambridge}
  \state{Massachusetts}
  \country{USA}
}

\author{Amir Yazdanbaksh}
\email{ayazdan@google.com}
\affiliation{%
  \institution{Google Research, Brain Team}
  \city{Mountain View}
  \state{California}
  \country{USA}}

\author{Shvetank Prakash}
\email{sprakash@g.harvard.edu}
\affiliation{%
  \institution{Harvard University}
  \city{Cambridge}
  \state{Massachusetts}
  \country{USA}}

\author{Jason Jabbour}
\email{jasonjabbour@g.harvard.edu}
\affiliation{%
 \institution{Harvard University}
 \city{Cambridge}
 \state{Massachusetts}
 \country{USA}}

 \author{Ikechukwu Uchendu}
 \email{iuchendu@g.harvard.edu}
\affiliation{%
 \institution{Harvard University}
 \city{Cambridge}
 \state{Massachusetts}
 \country{USA}}

 \author{Susobhan Ghosh}
 \email{susobhan_ghosh@g.harvard.edu}
\affiliation{%
 \institution{Harvard University}
 \city{Cambridge}
 \state{Massachusetts}
 \country{USA}}

\author{Behzad Boroujerdian}
\email{behzadboro@utexas.edu}
\affiliation{%
\institution{UT Austin/Harvard University}
\city{Cambridge}
\state{Massachusetts}
\country{USA}}

\author{Daniel Richins}
\email{drichins@utexas.edu}
\affiliation{%
\institution{UT Austin}
\city{Austin}
\state{Texas}
\country{USA}}

\author{Devashree Tripathy}
\email{devashreetripathy@iitbbs.ac.in}
\affiliation{%
\institution{IIT Bhubaneswar/Harvard University}
\city{Bhubaneswar}
\state{Odisha}
\country{India}}
 
\author{Aleksandra Faust}
\email{faust@google.com}
\affiliation{%
  \institution{Google Research, Brain Team}
  \city{Mountain View}
  \state{California}
  \country{USA}}
 
\author{Vijay Janapa Reddi}
\email{vj@eecs.harvard.edu}
\affiliation{%
  \institution{Harvard University}
  \city{Cambridge}
  \state{Massachusetts}
  \country{USA}
}

\renewcommand{\shortauthors}{Krishnan et al.}

\begin{abstract}
Machine learning (ML) has become a prevalent
approach to tame the complexity of design space exploration
for domain-specific architectures.
While appealing, using ML for design space exploration poses several challenges.
First, it is not straightforward to identify the most suitable algorithm from an ever-increasing pool of ML methods.
Second, assessing the trade-offs between performance and sample efficiency across these methods is inconclusive.
Finally, the lack of a holistic framework for fair, reproducible, and objective comparison across these methods hinders the progress of adopting ML-aided architecture design space exploration and impedes creating repeatable artifacts.
To mitigate these challenges, we introduce \ours, an open-source gymnasium and easy-to-extend framework that connects a diverse range of search algorithms to architecture simulators.
To demonstrate its utility, we evaluate \ours across multiple vanilla and domain-specific search algorithms in the design of a custom memory controller, deep neural network accelerators, and a custom SoC for AR/VR workloads, collectively encompassing over 21K experiments.
The results suggest that with an unlimited number of samples, ML algorithms are equally favorable to meet the user-defined target specification if its hyperparameters are tuned thoroughly; no one solution is necessarily better than another (e.g., reinforcement learning vs. Bayesian methods).
We coin the term ``\textit{hyperparameter lottery}'' to describe the relatively probable chance for a search algorithm to find an optimal design provided meticulously selected hyperparameters.
Additionally, the ease of data collection and aggregation in \ours facilitates research in ML-aided architecture design space exploration.
As a case study, we show this advantage by developing a proxy cost model with an RMSE of 0.61$\%$ that offers a 2,000-fold reduction in simulation time.
Code and data for \ours is available at \href{https://bit.ly/ArchGym}{https://bit.ly/ArchGym}. 
\end{abstract}






\maketitle

\section{Introduction}
\label{sec:intro}
Hardware customization~\cite{tpu,jouppi2023tpu,brainwave,nvidia:h100,chen2016eyeriss,yazdanbakhsh2021evaluation,tensor-cores,Tambe_isscc2021,intel-vnni} has played a pivotal role in realizing the potential of machine learning in various applications~\cite{alpha-fold,drug-discovery,rl-nuclear-fusion,rl-chip-design,rl-robotics,wang2021nerf,autopilot}. The stagnation of Moore's law and the increasing demand for compute efficiency~\cite{compute_efficiency,palm,gpt3,rasley2020deepspeed} have propelled the field towards pursuing extreme domain-specific customization.
While intriguing, this direction poses several challenges.
The immense number of design parameters across the compute stack leads to a combinatorial explosion of the search space~\cite{fast,yazdanbakhsh2021evaluation,mindmapping}.
Within this space, numerous infeasible design points further complicate optimization~\cite{mindmapping,kumar2021data}. 
Additionally, the diversity of the application landscape and the unique characteristics of the search space across the compute stack challenge the performance of conventional optimization methods.
To address these challenges, both industry~\cite{rl-chip-design,nvidia-rl,dsoi} and academia~\cite{mindmapping,vaesa,architect,yazdanbakhsh2021evaluation,granite,gamma,kao2020confuciux} have turned towards ML-driven optimization to meet stringent domain-specific requirements. 
Although prior work has demonstrated the benefits of ML in design optimization, the lack of reproducible baselines hinders fair and objective comparison across different methods.% and complicates their deployment.

First, \textit{selecting the most suitable algorithms and gauging the role of hyperparameters and their efficacy is still inconclusive}.
There are a wide range of ML/heuristic methods, from random walker~\cite{rw} to reinforcement learning (RL)~\cite{rl}, that can be employed for design space exploration (DSE).
%
% Sri's version to write the same in non-critical/controversial way. I agree with you. Can't see seen as criticism of their work 
For example, recent work has applied Bayesian~\cite{ReagenHAGWWB17,milad-bo}, data-driven offline~\cite{yazdanbakhsh2021evaluation}, and RL~\cite{kao2020confuciux} optimization methods for architecture parameter exploration of Deep Neural Network (DNN) accelerators.
% suggested not to use due (because of is better)
While these methods have shown noticeable performance improvement over their choice of baselines, it is not evident whether the improvements are because of the choice of optimization algorithms or hyperparameters. 
To ensure reproducibility and facilitate widespread adoption of ML-aided architecture design space exploration, it is imperative to outline a systematic benchmarking methodology.
%
%\ayazdan{this sentence could be interpreted negatively. I know we used it before. Let's talk I could be wrong.}
%
%For instance, prior work has applied Bayesian~\cite{ReagenHAGWWB17,milad-bo}, data-driven offline~\cite{yazdanbakhsh2021evaluation}, and RL~\cite{kao2020confuciux} optimization methods for architecture parameter exploration of Deep Neural Network (DNN) accelerators.
%
%These works compare their algorithm against the others as baselines and all show that their algorithm has better and noticeable performance improvements over the others.
%
%Given this disparity in comparison to different baselines, it is not evident whether the reported improvements are owed to the choice of optimization algorithms or hyperparameters.
%
%Understanding why and to what we owe the improvements is critical for the widespread adoption of ML as a tool for architecture design space exploration.
%
%And as such, it is imperative to outline how to perform apples-to-apples benchmarking to reproducibly compare search algorithms across various architectural exploration problems and solutions.

Second, while simulators have been the backbone of architectural innovations, \textit{there is an emerging need to address the trade-offs between  accuracy, speed, and cost in architecture exploration}. 
The accuracy and performance estimation speed widely varies from one simulator to another, depending on the underlying modeling details (e.g. cycle-accurate~\cite{sniper,gem5} $\rightarrow$ transaction-level simulator~\cite{dramsys,simics} $\rightarrow$ analytical model~\cite{farsi,parashar2019timeloop,kwon2020maestro,arafa2019ppt,wu2021sparseloop} $\rightarrow$ ML-based proxy models~\cite{tpu-ml-model,granite,hashemi2018learning, mendis2019ithemal}).
While analytical or ML-based proxy models are nimble by virtue of discarding low-level details, they generally suffer from high prediction error.
Also, due to commercial licensing, there can be a strict limits on the number of samples collected from a simulator~\cite{hardware-licensing}.
Overall, these constraints exhibit distinct performance vs. sample efficiency trade-offs, affecting the choice of optimization algorithm for architecture exploration.
Therefore, it is challenging to delineate \textit{how to  systematically compare the effectiveness of various ML algorithms under these constraints.}

\begin{figure}[t]
  \centering
  \includegraphics[width=0.9\columnwidth]{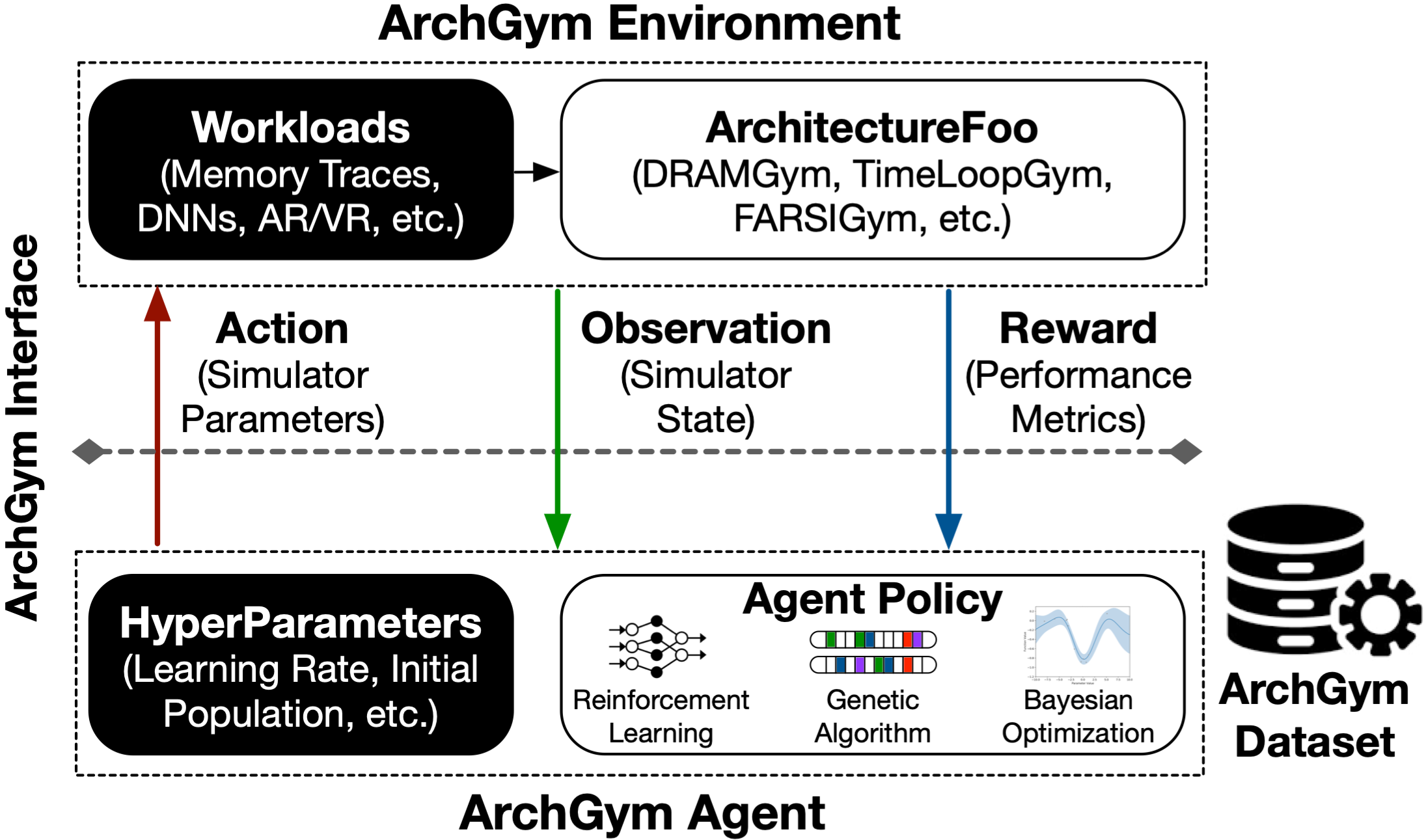}
  \caption{\rev{\ours comprises two main components: the `ArchitectureFoo' environment and the `Agent'. ArchitectureFoo encapsulates the cost model, which can be a simulator (e.g., DRAMSys~\cite{dramsys}), a roofline model (e.g., FARSI~\cite{farsi}), an analytical model (e.g., MASTERO~\cite{kwon2020maestro}), or even real hardware. Similarly, the second component, Agent, is an abstraction of a policy and hyperparameters (see Section~\ref{sec:arch-gym-interface}). With a standardized interface that connects these two components, \ours provides a unified framework for evaluating different machine learning-based search algorithms fairly while also saving the exploration data as the \ours Dataset. By using \ours, researchers and practitioners can compare and evaluate the performance of different algorithms in a consistent and systematic manner.}}
  \label{fig:arch-gym-design} 
  \vspace{-1.7em}
\end{figure}

Lastly, \textit{rendering the outcome of DSEs into meaningful artifacts such as datasets is critical for drawing insights about the design space}.
It is commonly known that the landscape of ML algorithms is rapidly evolving and some ML algorithms~\cite{offline-rl} need data to be useful.
Solely in the RL domain, we have witnessed the emergence of several algorithmic formulations (\eg~PPO~\cite{ppo}, SAC~\cite{sac}, DQN~\cite{dqn}, DDPG~\cite{ddpg}) solving a variety of problems. 
In parallel, recent efforts have employed offline RL~\cite{offline-rl} methods to amortize the cost of data collection. In this rapidly evolving ecosystem, it is consequential to ensure \textit{how to amortize the overhead of search algorithms for architecture exploration.}
%
%It is not apparent, nor systematically studied, whether we can reuse the collected data from one search algorithm to another.
It is not apparent, nor systematically studied how to leverage exploration data while being agnostic to the underlying search algorithm.

To alleviate these challenges, we introduce \ours (See \Fig{fig:arch-gym-design}), an open-source gymnasium to analyze and evaluate various ML-driven methods for design optimization.
\ours reinforces using the same interface between search algorithms and performance models (e.g. architecture simulator or proxy cost models), enabling effective mapping of variety of search algorithms.
%
% Amir -> the format of exchanging informat? The sentence by itself does not make sense? how come these critiera enable fairness???
% \ayazdan{In addition, for a fair comparisons, we propose two criteria: (i) the search algorithms must interact with the performance models (e.g. architecture simulator or proxy cost models) through the same interface; 
%
% (ii) the exchange of information, such as state and reward, between the search algorithms and performance models remains the same.}
% %
% Exhibiting these criteria enables \ours to effectively map different search algorithms in a holistic and unified manner.
%
This interface also forms the scaffold to develop baselines for comparison and benchmarking of search algorithms, as they continue to evolve and grow. 
Furthermore, \ours provides an infrastructure to collect and share datasets in a reproducible and accessible manner, which organically advances the understanding of the underlying design spaces and improves the status quo search algorithms. 

We perform more than 21,600 experiments corresponding to around 1.5 billion simulations across four architectural design space exploration problems, (1) DRAM memory controller, (2) DNN accelerator, (3) SoC design, and (4) DNN mapping.
We use five commonly used search algorithms and comprehensively sweep their associated hyperparameters.
Our evaluation shows that there are significant variations in the final performance of these algorithms. 
\begin{table}[t]
  \caption{Prior works using ML for optimizing architectural components.} %However, reproducing these works is challenging due to the lack of a common evaluation infrastructure and hyperparameter details.}
  \resizebox{1.0\columnwidth}{!}{
\begin{tabular}{|c|c|c|}
\hline
{\color{black} \textbf{Prior Work}}                    & {\color{black} \textbf{ML-Method}}                                       & {\color{black} \textbf{Architecture Component}}                                      \\ \hline
{\color{black} DRL-NOCS~\cite{drl-nocs}}                               & {\color{black} Reinforcement Learning}                                   & {\color{black} Network-on-Chips}                                                     \\ \hline
{\color{black} GAMMA~\cite{gamma}}                                  & {\color{black} Genetic Algorithm}                                        & {\color{black} ML Accelerator Mapping}                                               \\ \hline
{\color{black} PRIME~\cite{kumar2021data}}                                  & {\color{black} Data-Driven Offline Learning}                                         & {\color{black} ML Accelerator Datapath}                                              \\ \hline
{\color{black} Reagen. et. al.~\cite{ReagenHAGWWB17}}                        & {\color{black} Bayesian Optimization}                                    & {\color{black} NN HW-SW Co-Design}                                                   \\ \hline
{\color{black} FAST~\cite{fast}}                                   & {\color{black} Linear Combinatorial Swarms}                              & {\color{black} NN HW-SW Co-Design}                                                   \\ \hline
{\color{black} Ipek. et. al.~\cite{rl-dram}}                          & {\color{black} Reinforcement Learning}                                   & {\color{black} DRAM Memory Controller}                                               \\ \hline
{\color{black} Zhang. et. al.~\cite{aco-circuit-design}}                         & {\color{black} Ant Colony Optimization}                                  & {\color{black} Circuit Parameters}                                                   \\ \hline
{\color{black} Compiler Gym~\cite{compiler-gym}}                           & {\color{black} Reinforcement Learning}                                   & {\color{black} Compiler Optimization}                                                \\ \hline\hline
\multicolumn{1}{|l|}{{\color{black} \textbf{\ours (This Work)}}} & \multicolumn{1}{l|}{{\color{black} \textbf{ML * (BO, GA, RL, ACO ...)}}} & \multicolumn{1}{l|}{{\color{black} \textbf{Architecture* (DRAM, SoC, Mapping etc)}}} \\ \hline
\end{tabular}
    }
    \label{tab:related-work}
\end{table}
For instance, we observe a statistical spread in the performance of different search algorithms of up to 90\%, 20\%, and 40\% for DRAM memory controller, DNN Accelerator, and SoC design, respectively.\footnote{We measure the statistical spread by reporting the interquartile range.}
Including outliers, each algorithm yields at least \textit{one configuration} that achieves the \textit{best} objective across different design spaces.

These observed variations is the results are primarily a consequence of hyperparameter selection, as yet an open research problem~\cite{hyperparameter-tuning}.
The choice of optimal hyperparameter values depends on the characteristics of the ML algorithm as well as the underlying domain.
% more resources, especially to domain.
However, commonly used hyperparameter tuning techniques~\cite{randomwalker-hopt,bayesopt-hopt,hypopt-book} introduce another layer of complexity. 
That is, identifying the optimal hyperparameters for architecture DSE remains non-trivial, an improbable task akin to winning a lottery, requiring significant amount of resources.
This ``\textit{hyperparameter lottery}'' describes the comparable chance of an algorithm to attain an optimal solution.\footnote{We refer to a design as optimal as long as it meets all user-defined criteria for a target hardware, for example latency $<L$.}
Finally, in contrast to common wisdom, our analyses suggest that the evaluated search methods are equally favorable across different design space exploration problems. 
Below we summarize the main contributions of our work:
\begin{itempacked}
\item We design and open source the \ours framework for ML-aided architecture design space exploration, enabling systematic evaluation and objective comparison of search algorithms.% for exploration.
\item Leveraging this framework, we show that, contrary to common wisdom, the evaluated search algorithms are all equally favorable for architecture design space exploration, no one algorithm (e.g. RL or Bayesian methods or GA) is necessarily more promising.
%
% Amir -> don't write long sentences, always break and short.
\item We argue that to fairly compare ML algorithms, it is crucial to take into account the cost of hyperparameter optimization, such as access to hardware simulator samples. Without proper evaluation metrics, the effectiveness of search algorithms can be misleading to realize the potential of ML-aided design.
\item We release a set of curated datasets that are useful for building high-fidelity proxy cost models. Such proxy cost models are often orders of magnitude faster compared to conventional cycle-accurate simulators, mitigating the trade-off between speed and accuracy in architecture exploration.   
\item Building off the intuition that increasing dataset size improves accuracy, we show that adding diversity, enabled by \ours, can reduce the average root mean square error by up to 42$\times$.
\end{itempacked}

\section{Background and Related Work}
Though ML can be used for many classes of optimization problems, in this paper, we center our study to architecture design space exploration problems.
Architecture DSE corresponds to a class of problems that uses search algorithms to navigate the architecture parameter design space.
This generally forms a prohibitively large search space, and as a result an intractable problem for manual search.
Hence, architects commonly employ heuristics or ML-aided search algorithms to navigate the space in the pursue of efficient designs. 
One of the common metrics to asses the efficiency of search algorithms is the number of requisite samples before reaching an optimal solution. 
The search algorithm iteratively suggests parameter values for a given workload (or set of workloads).
The fitness (i.e., \emph{how good a particular parameter selection is}) of these selections is determined by a cost model. 
For architecture DSE, this cost model can be a time- and resource-consuming cycle-accurate simulators or a relatively fast inaccurate analytical models.
\begin{figure}[t!]
  \centering
  \includegraphics[width=\columnwidth]{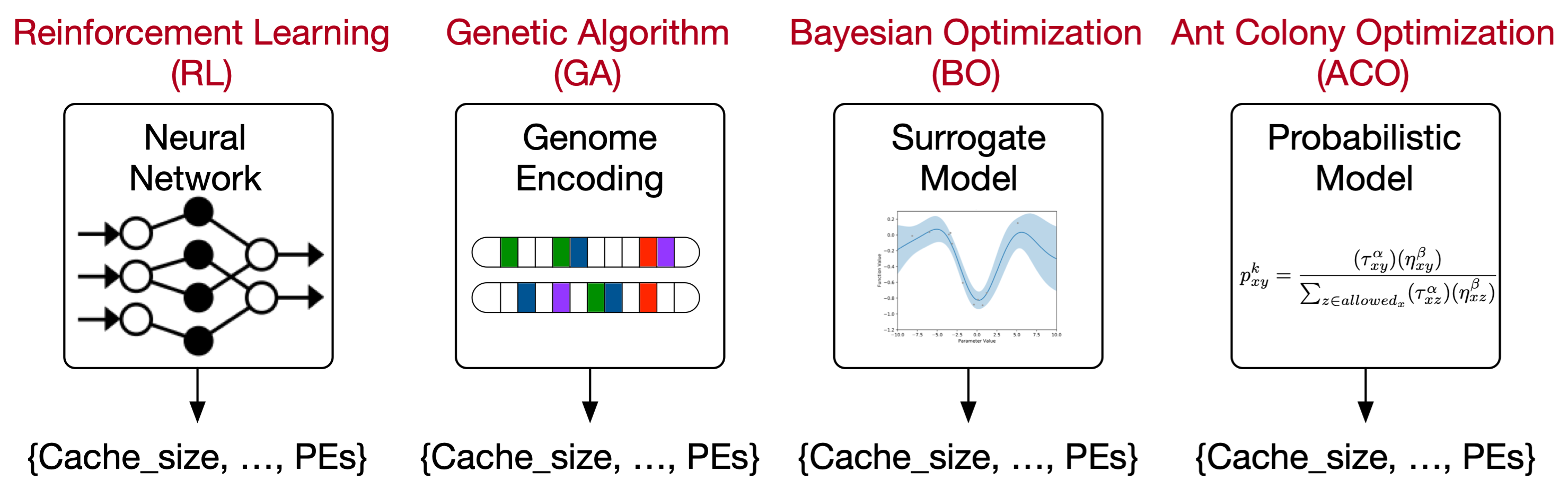}
  \caption{Agent and its policy are used to determine optimal parameter selection. For example, in an RL agent, the policy is typically a neural network. In Genetic algorithms (GA), the policy is a genome. In BO, the policy is a surrogate model; in ACO, the policy is a probabilistic model. The policies in each of these agents determine the parameter selection. For example, in micro-architectural resource allocation problems, parameters can be any micro-architectural parameters such as cache size or PE counts.} 
  \label{fig:agent} 
  \vspace{-0.6cm}
\end{figure}

While an exhaustive search may be feasible when number of parameters is modest, such approach is not practical even in automated DSE frameworks~\cite{de2020automated, snider2001spacewalker,hourani2009automated}.
A growing body of work has used analytical models~\cite{karkhanis2004first, noonburg1994theoretical}, sampling techniques~\cite{wunderlich2003smarts, sherwood2002automatically}, and statistical simulation~\cite{rao2002hlspower, oskin2000hls, eeckhout2004control} to navigate large search spaces.
However, the continued increase in configurable architecture parameters~\cite{fast,yazdanbakhsh2021evaluation} is expanding the design space~\cite{kang2010approach}, straining conventional search methods.

To mitigate this, ML algorithms have been widely employed in searching architectural parameters with inspiring results as some are highlighted Table~\ref{tab:related-work}.
Bayesian optimization (BO)~\cite{ReagenHAGWWB17,bo2,bo3} is a popular technique for tuning architectural parameters.
However, since the time complexity of BO is cubic to the number of samples~\cite{bayesopt-cubic-1,bayesopt-cubic-2}, it is yet to be seen if BO can successfully converge on extremely large design spaces~\cite{kao2020confuciux,fast,kumar2021data}.
Alternatively, single agent RL is used for data path and mapping optimization for DNN accelerators~\cite{kao2020confuciux}, identifying quantization levels~\cite{elthakeb2020releq}, and commercial chip floor planning~\cite{rl-chip-design}.
% Confuciux~\cite{kao2020confuciux} uses RL for data path and mapping choices of DNN accelerators. 
% %
% ReLeQ~\cite{elthakeb2020releq} applies RL to identify the optimal setting for the quantization levels of deep neural networks.
% %
% Furthermore, CircuitTraining~\cite{rl-chip-design} employs RL algorithm for chip floor planning in a commercial setting.
%
% These works generally expose a large design space to a single RL agent.
%
However, it is commonly-known that RL algorithms are sample inefficient~\cite{rl-sample-efficiency-1,rl-sample-efficiency-2} and susceptible to variations in hyperparameter initialization~\cite{hyp-rl}.
It is evident that both the machine learning and hardware communities could benefit from developing stronger introspection skills in order to create more resilient and dependable RL optimization algorithms.
%
% It is evident that both machine learning and hardware community lack introspection capabilities to produce robust and reliable RL optimization algorithms.
%
%Though ML can be used for many classes of optimization problems, in this paper, we center our study to architecture design space exploration problems.  
%

%\input{Arch-Gym/tex/dse_problem}
\section{Design and Implementation}

\begin{table*}[ht!]
\caption{The similarity between various ML-based search algorithms. \ours uses these similarities to integrate the different algorithms into a standard, unified interface.}
\label{tab:common-interfaces}
\resizebox{\linewidth}{!}{
\begin{tabular}{|c|cccc|c|}
\hline
\multirow{2}{*}{\textbf{Q\textit{n}}} & 
\multicolumn{4}{c|}{\textbf{Search Algorithms}}                    & 
\multirow{2}{*}{\textbf{Requirements}}                                   
                     \\ \cline{2-5}
                                   & \multicolumn{1}{c|}{\textbf{\begin{tabular}[c]{@{}c@{}}Reinforcement\\ Learning\end{tabular}}}   & \multicolumn{1}{c|}{\textbf{\begin{tabular}[c]{@{}c@{}}Bayesian\\ Optimization\end{tabular}}}                & \multicolumn{1}{c|}{\textbf{\begin{tabular}[c]{@{}c@{}}Ant-Colony\\ Optimization\end{tabular}}}                            & \textbf{\begin{tabular}[c]{@{}c@{}}Genetic\\ Algorithms\end{tabular}}                                                                                                                                                          &                                                          \\ \hline
\textbf{Q1}                        & \multicolumn{1}{c|}{\begin{tabular}[c]{@{}c@{}}Policy will determine\\ the actions\end{tabular}} & \multicolumn{1}{c|}{\begin{tabular}[c]{@{}c@{}}Surrogate model\\ will determine the \\ actions\end{tabular}} & \multicolumn{1}{c|}{\begin{tabular}[c]{@{}c@{}}Pheromones will determine\\ the actions\end{tabular}}                       & \begin{tabular}[c]{@{}c@{}}Genome will determine\\ the actions\end{tabular}                                & \begin{tabular}[c]{@{}c@{}}Reward/Fitness\\ is needed\end{tabular}                                                       \\ \hline
\textbf{Q2}                        & \multicolumn{1}{c|}{\begin{tabular}[c]{@{}c@{}}Reward is used\\ as feedback\end{tabular}}        & \multicolumn{1}{c|}{\begin{tabular}[c]{@{}c@{}}Fitness is used as\\ feedback\end{tabular}}                   & \multicolumn{1}{c|}{\begin{tabular}[c]{@{}c@{}}Fitness value will determine\\  pheromone concentration\end{tabular}}       & \begin{tabular}[c]{@{}c@{}}Fitness value will determine \\ which candidates need to reproduce\end{tabular} & \begin{tabular}[c]{@{}c@{}}State of the simulator/observation is\\ needed to understand how good/bad it is.\end{tabular}  \\ \hline
\textbf{Q3}                        & \multicolumn{1}{c|}{Hyperparameters in RL algorithm}                                             & \multicolumn{1}{c|}{\begin{tabular}[c]{@{}c@{}}Acquisition\\ Function\end{tabular}}                          & \multicolumn{1}{c|}{\begin{tabular}[c]{@{}c@{}}Stochastic model will determine \\ when to explore vs exploit\end{tabular}} & Mutation/Cross-over operation                                                                              & Intrinsic to Agent                                                                                                        \\ \hline
\end{tabular}}
\end{table*}

Our primary objectives in \ours is to apply ML approaches for architecture design space exploration, fairly compare their performance, and provide an infrastructure for collecting exploration datasets.
Additionally, we aim to understand the relevant trade-offs associated with different ML algorithms.

% To achieve these objectives, we introduce \ours.

%
% This section discusses various design components in \ours.
%
%In addition, we provide the intuitions behind constructing a standardized interface for benchmarking ML-based search algorithms. 
%
\ours consists of three main components, namely \emph{Gym Environment}, \emph{Agents}, and \emph{Interface Signals}.
Figure~\ref{fig:arch-gym-design} outlines a high-level view of these components in \ours.
Our proposed search framework is designed to be modular and flexible to exchange the architectural problem under study as well as ML agents in a straightforward manner. 
To achieve this, we wrap each architectural component into an \ours environment.
The architectural cost model can be a cycle-accurate simulator, analytical model, an ML-based cost model, or alternatively silicon hardware. 
There are bi-directional interface signals from the \ours environment to the agents.
Each agent, irrespective of its type (e.g., Bayesian optimization, reinforcement learning, ant colony optimization), uses the exact same interface signals to interact with \ours environments.

\subsection{Environment}
\label{sec:env}
Each environment is an encapsulation of the architecture cost model along with the target workload(s).
The architecture cost model determines the cost of running the workload, given a set of architecture parameters. 
For example, the cost can be latency, throughput, area, energy, or any other combinations of user-defined performance metrics. 
\Fig{fig:arch-gym-design} outlines various components in the environment. We expound each part in the following.

\niparagraph{Architecture cost model.}
Depending upon the architecture under study, `ArchitectureFoo' (see  \Fig{fig:arch-gym-design}) can be replaced with representative architecture cost model.
This is a placeholder for an architecture cost model in which a user intends to apply ML methods for design space exploration. 
For instance, if the user wants to use ML for architecture design space exploration of a memory controller, the ArchitectureFoo would be replaced by \textit{DRAMGym}, which encapsulates the DRAM architecture cost model.
Similarly, if the user wants to employ ML for design space exploration of DNN accelerators, ArchitectureFoo would be replaced by \textit{TimeloopGym} which encapsulates Timeloop~\cite{parashar2019timeloop}.
Finally, for a complex SoC for AR/VR workloads, we can encapsulate FARSI~\cite{farsi} as a \textit{FARSIGym} to provide the SoC cost model.

\niparagraph{Target Workload(s).}
Workloads are the integral components in \ours.
Each workload representation can be diverse and vary significantly depending upon the component the user intends to optimize.
For instance, in the case of optimizing memory controllers, the workload could be memory access traces of a particular application.
Likewise, for DNNs, the workload can be represented as a graph or information about various layers.
Similarly, an AR/VR workload can be represented as a task graph where each task can be mapped to different IPs in an SoC. 
\subsection{Agent}
\label{sec:agents}
We define `Agent'  as an encapsulation of the machine learning algorithm used for search.
An ML algorithm consists of `hyperparameters' and a guiding `policy'.
The hyperparameter is intrinsic to an algorithm which can significantly influences its performance.
A policy, on the other hand, determines how the agent selects a parameter iteratively to optimize its target objective.
To further reinforce this abstraction, we seed our infrastructure with five agents from recently developed search algorithms, namely, Ant-Colony Optimization~\cite{aco}, Bayesian Optimization~\cite{bo}, Genetic Algorithms~\cite{ga}, Random Walker agent~\cite{rw}, and Reinforcement Learning~\cite{rl} to solve the same set of architectural design space exploration problems. 
Figure~\ref{fig:agent} demonstrates the four agents that we develop and integrate in \ours.
The random walker algorithm~\cite{rw}, which is not shown in this figure, is merely a random search with a random number generator as its policy.

As shown in Table~\ref{tab:related-work}, these algorithms are commonly used in hardware design at different abstraction levels.
For example, BO has been used in efficient accelerator design space exploration~\cite{ReagenHAGWWB17} and in selecting the best coherency interfaces for hardware accelerators on SoCs~\cite{bo2}.
RL has been used for DNN architecture design~\cite{kao2020confuciux}, chip floor planning~\cite{rl-chip-design}, and exploring router-less NoC designs~\cite{lin2019optimizing}.
GA has been used in design space exploration of heterogeneous multi-processor embedded systems~\cite{muttillo2020openmp} and behavioral IPs ~\cite{schafer2017parallel}.
ACO has been applied to solving long-standing compiler optimization problems~\cite{shobaki2022register} and has been used in high-level synthesis design space exploration~\cite{gao2021effective}. 
\subsection{Interface}
\label{sec:interface}
An Agent's role in architecture design space exploration is to output a set of parameter selections according to its policy.
This parameter selection acts as an input signal to the \ours environment.
Likewise, the \ours's environment outputs the state information of the architecture cost model and a feedback signal back to the agent.
We need interface signals to facilitate these communications between the agent and the \oursnospace.
Three main signals are needed: \texttt{action}, \texttt{observation}, \texttt{reward} (\Fig{fig:arch-gym-design}).
They all use these signals to optimize their target objective regardless of the agent type.

\begin{table*}[!ht]
  \centering
  \caption{Summary of the Gym environments in experimental setup. Objective is to optimize reward via optimal actions.}
\resizebox{\linewidth}{!}{
    \begin{tabular}{|c||c|c|c|c||c|}
    \hline
    \textbf{\begin{tabular}[c]{@{}c@{}}Gym \\ Environment\end{tabular}} & 
    \textbf{\begin{tabular}[c]{@{}c@{}}Simulator \end{tabular}} & 
    \textbf{\begin{tabular}[c]{@{}c@{}}Workload \end{tabular}} & 
    \textbf{\begin{tabular}[c]{@{}c@{}}Action \end{tabular}}  &
    \textbf{\begin{tabular}[c]{@{}c@{}}Observation \end{tabular}}  &
    \textbf{\begin{tabular}[c]{@{}c@{}}Reward \end{tabular}}  \\
    
    \hline{{\textsc{DRAMGym}}}  & {DRAMSys~\cite{dramsys}} & {\shortstack{Memory Trace \\ (i.e., Streaming Access Pattern, Random Access Pattern)}} & {\shortstack{Memory Controller Parameters \\ (i.e, RefreshPolicy, RequestBufferSize, etc.)}} & {\texttt{<latency, power, energy>}} & {$r_{x} = \frac{X_{\text{target}}}{|X_{\text{target}}-X_{\text{obs}}|}$} \\
   
    \hline{{\textsc{TimeloopGym}}}  & {Timeloop~\cite{parashar2019timeloop}} & {\shortstack{Convolutional Neural Network \\ (i.e., AlexNet, MobileNet, ResNet-50)}} & {\shortstack{Accelerator Parameters \\ (i.e., NumPEs, WeightsSPad\_BlockSize, etc.)}} & {\texttt{<latency, energy, area>}} &{$r_{x} = \frac{X_{\text{target}}}{|X_{\text{target}}-X_{\text{obs}}|}$}\\
    
    \hline{{\textsc{FARSIGym}}}  & {FARSI~\cite{farsi}} & {\shortstack{AR/VR Audio \& Image Processing Task Dependency Graph \\ (i.e., Audio Decoder, Edge Detection)}} & {\shortstack{System On-Chip Parameters \\ (i.e., PE\_Type, NoC\_BusWidth, etc.)}} & {\texttt{<power, performance, area>}} & \shortstack{{$distance\ to\ budget = \sum_m \alpha*\frac{(D_m - B_m)}{B_m}$} \\ {$m \in \{ Performance, Power, Area\} \nonumber$}}\\

    \hline{{\textsc{MaestroGym}}}  & {Maestro~\cite{kwon2020maestro}} & {\shortstack{DNN Workloads \\ (i.e., ResNet18, MobileNet, etc.)}} & {\shortstack{DNN Mapping \\ (i.e., L1 and L2 mapping, etc.)}} & {\texttt{<runtime, throughput, energy, area>}} & 
    {$r_{x} = \frac{1}{X_{\text{target}}}$}\\
    
    \hline\end{tabular}}\label{tab:gym_envs}
\label{tab:gymenv_summary}
\end{table*}

\ours uses the OpenAI gym interface to expose the parameters to all machine learning algorithms.
The \texttt{action} is used as the interface to relay the agents actions to the environment. Likewise, the \texttt{observation} is used as the interface to relay the state information of the environment back the agent. Additionally, the \texttt{reward} is the feedback signal for the agent's architecture parameter selection.
In the case of RL, this signal is called the reward signal. Likewise, this reward signal is called fitness in other agents, such as Bayesian optimization, ACO, and GA.
The agent uses this \texttt{reward} to fine-tune its policy to make better parameter selections in the future to optimize its target objective.

Since all agents interact with \ours environment through the same interface, we can additionally record each interaction. 
We call these interactions trajectories.
These trajectory recordings can be used to construct standardized datasets for training sample inefficient algorithms, offline algorithms~\cite{kumar2021data}, or constructing cost models for architectural simulators (See Section~\ref{sec:results}).

\subsection{Dataset Generation}
% Data is the fuel of any ML-aided design, yet it is frequently overlooked.
%
One of the unique features of \ours is that it can be used to collect standardized datasets across all agents for the same architecture design space exploration task.
Using a standardized interface (Section~\ref{sec:interface}), \ours can log the information exchanges between all agents and the environment into popular dataset exchange formats like TFDS~\cite{paper2021tensorflow}, RLDS~\cite{ramos2021rlds}. 
Over time, accruing these datasets for similar architecture design space explorations enables new variants of data-driven offline learning algorithms~\cite{yazdanbakhsh2021evaluation}.
Furthermore, as we demonstrate in Section~\ref{sec:dataset}, the diversity in these datasets (from different agents) can create high-fidelity proxy cost models to replace the slow architectural cost models (e.g., cycle-accurate simulators).
We strongly believe a community-wide adoption of \ours methodology can create useful datasets for tackling fundamental problems surrounding data scarcity from the architecture cost model and can significantly speed up architecture design space exploration. 
\begin{figure*}[!t]
\centering
  \includegraphics[width=0.96\textwidth,keepaspectratio]{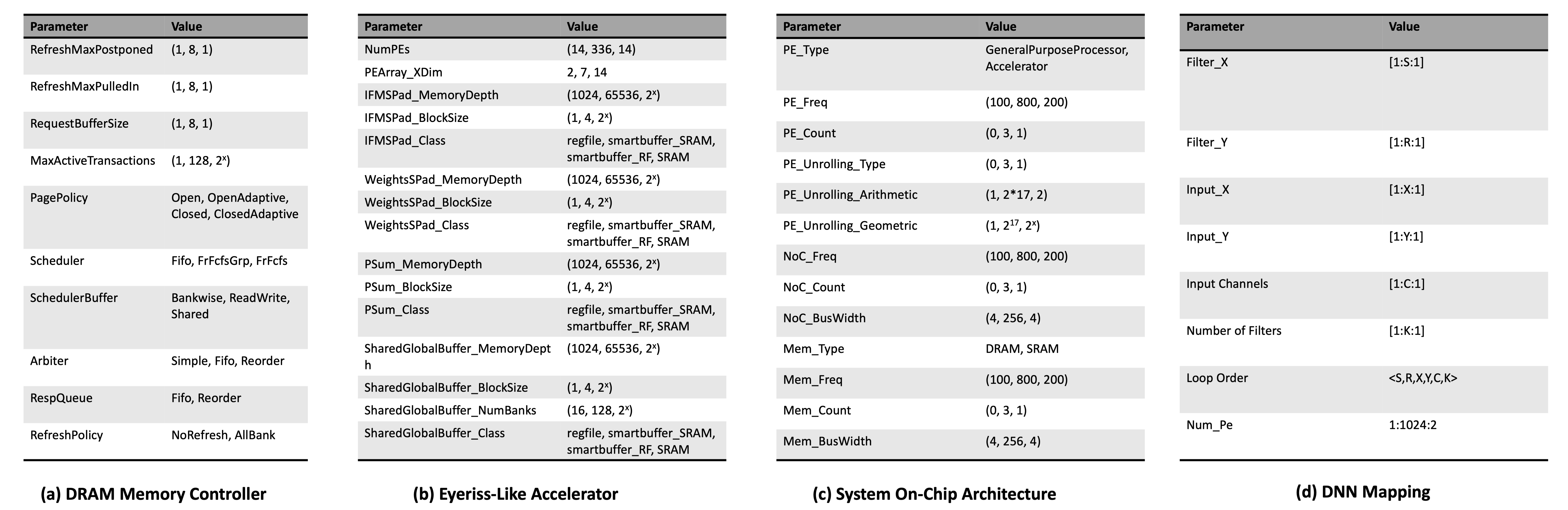}
  \caption{Architecture DSE problem from (a) component-level , (b) accelerator level, and (c) SoC level, and (d) Mapping. A mixture of numerical and categorical parameters are learned. Numerical parameters are specified in tuple format: (min, max, step). The total search space for DRAMGym, TimeloopGym, FARSIGym, and MASTEROGym are 1.9e7, 2e14, and 1.6e17, 1e24 (for VGG16 second layer) respectively.}
  \label{fig:arch_dse_problem}
\end{figure*}
\section{Integration of New Algorithms}
\label{sec:arch-gym-interface}
In this section, we provide an intuitive explanation of the design and implementation of \ours for ML-aided design space exploration. This approach allows us to better understand the similarities between different ML search algorithms and serves as a foundation for adding novel ML-based search algorithms to \ours.

The goal of any intelligent agent is to determine the optimal parameter selection to maximize (or minimize) the objective. During the agents' training process (optimization phase), it has to fine-tune its policy. To achieve these goals, each agent answers the following questions:  \emph{\textbf{(Q1)}: How does an agent determine a particular parameter (action)?}; \emph{\textbf{(Q2)} How does an agent use feedback for selecting a particular parameter and fine-tune the policy?}; \emph{\textbf{(Q3)} How does the agent balance between exploration and exploitation?} We tabulate the answer to these questions for the four agents shown in Table~\ref{tab:common-interfaces}.

Given that each agent approaches the same questions differently, we distill the information exchanged by all these agents to answer \textbf{Q1}, \textbf{Q2}, and \textbf{Q3} to standardize the interface. Such standardization of interfaces provides baselines to objectively compare all the agents in a fair and reproducible manner.

\niparagraph{(Q1) How does each agent select a parameter?}
In the Q1 scenario, each agent has a policy that allows it to take intelligent parameter selection. 
As shown in \Fig{fig:agent}, RL uses the neural network policy to determine the action. In BO, the agent's surrogate model determines the actions. 
Likewise, in ACO, the agent has a simple probabilistic model as a function of pheromone concentration, guiding the ant agent to take a particular action. Lastly, in the case of GA, the genome of each population determines the actions. Hence, from an information exchange point of view, each agent outputs actions (parameters) while the policy being intrinsic to the agent. Moreover, since each search algorithm is iterative, the agent needs to output actions at each step/iteration.

\niparagraph{(Q2) How does the agent receive feedback after selecting a parameter?}
Once the agent selects a parameter, it is relayed to the environment.
The environment performs the simulation and outputs a feedback signal that the agent uses to fine-tune its policy.
In RL, the reward signal is a function of the optimization objective (e.g., minimizing latency, area, etc.).
In BO, ACO, and GA, it receives fitness metrics to evaluate the selected parameter set and fine-tune the policy accordingly.
For instance, in an ACO agent, the pheromone concentration is modulated by this fitness. Similarly, in BO, the surrogate model is refined.
Likewise, GA uses it to determine the fittest agent for natural selection, which will result in a new genome.
Hence, from an information exchange point of view, each agent receives a reward/fitness metric, which is consequently used to fine-tune the policy and take more intelligent parameter selection. 

\niparagraph{(Q3) How does the agent balance between exploration and exploitation?}
In a large parameter design space, an intelligent agent should ensure that it shouldn't get stuck at local maxima or minima. To learn this, it needs to explore other regions smartly, even though there is a higher chance that it would be sub-optimal. The efficacy of each agent in determining the optimal parameter set depends upon how the agent balances exploration and exploitation. For this purpose, each agent has a set of hyperparameters that influences how an agent performs exploration and exploitation. For example, in GA, there are two, namely probability of mutation and crossover, which allow the GA agent to explore.
Similarly, ACO has a probabilistic model which switches between selecting random actions or choosing a parameter set that has a higher pheromone concentration. When the ACO agent chooses a random action, it facilitates exploration. Likewise, in BO, there is an acquisition function~\cite{bo} that facilitates exploration. Lastly, in RL, many methods, such as adding noise to NN policy and regularization, etc., are employed to facilitate exploration in reinforcement learning. Hence, in summary, how agent balances exploration and exploitation is often an integral part of the agent. Therefore, each agent should expose its hyperparameters during the agent's initialization.

\begin{figure*}[!t]
  \centering
  \includegraphics[width=\linewidth]{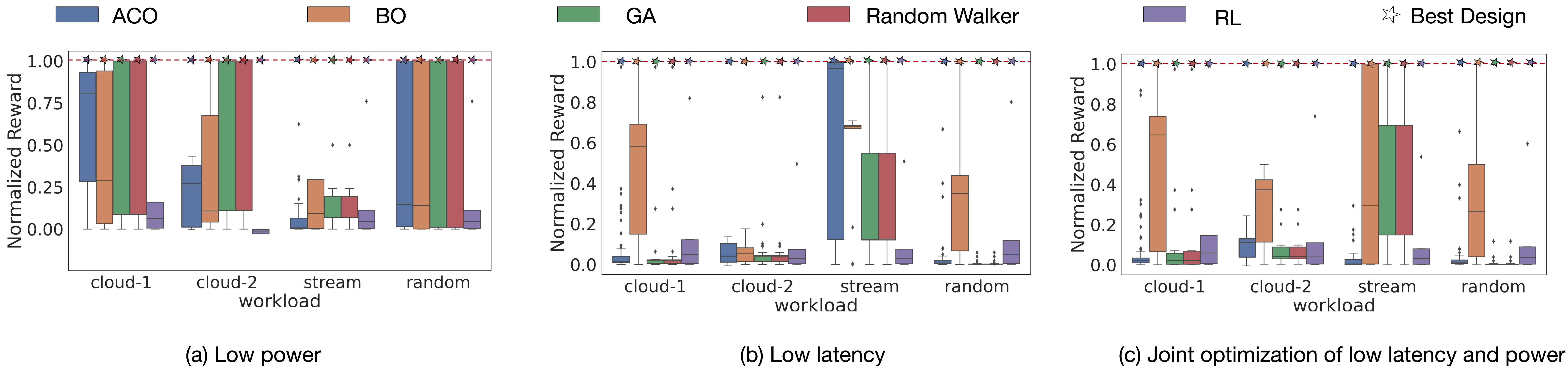}
  \caption{Hyperparameter lottery across different target objectives: low-power, low latency, and joint optimization of latency and power for DRAMGym environment. ACO, BO, GA, RL refers to ant colony optimization, Bayesian optimization, genetic algorithm, and reinforcement learning, respectively. The design that achieves the maximum reward is the best design and denoted by a star symbol.}
  \label{fig:dramsys-lottery} 
\end{figure*}
We adopt the OpenAI gym~\cite{openai} interface to integrate all the necessary information exchange to and from the agents. While OpenAI gym is limited to RL, Arch-Gym supports many other agents thanks to our observations that all agents can be systematically be broken down into Q1, Q2 and Q3. For example, in Q1, each agent uses the policy to determine a set of parameters. Furthermore, the gym environment provides a \texttt{step()} interface through which the agent's action (parameter) can be encapsulated. In the case of Q2, the \texttt{step()} also returns a feedback signal (e.g., reward/fitness) which the agents can use to fine-tune its policy. Lastly, in the case of Q3, each agent's hyperparameters are innate to the agent's initialization. Moreover, in certain RL algorithms (e.g., DDPG~\cite{ddpg}), noise is added to the parameters (i.e., action) to allow further exploration. These noise-induced parameters can be passed through the \texttt{step()}.

\section{Experimental Setup}
\label{sec:setup}
In this section we describe the simulator, workload, and agent implementations. 
Table~\ref{tab:gymenv_summary} summarizes the key aspects of each simulator, such as workloads, actions, observation, and rewards.
In Figure~\ref{fig:arch_dse_problem}, we provide the parameter set for each \ours environment and elaborate on our experimental setup.
Briefly, we use four environments to demonstrate the utility of \ours and their details (architecture, parameters, workloads).

\niparagraph{Simulators.} We developed gym environments for four simulators to demonstrate \ours's generalizability: DRAMGym uses DRAMSys~\cite{jung2015dramsys}, TimeloopGym uses Timeloop~\cite{parashar2019timeloop} for DNN accelerators, FARSIGym uses FARSI~\cite{farsi} for complex SoCs, and MaestroGym uses Maestro~\cite{kwon2020maestro} for DNN mapping as the simulator. 

\niparagraph{Workloads.} For exploring DRAM memory controller designs, we use the memory traces provided within DRAMSys~\cite{dramsys}. Likewise, for evaluating several candidate SoC designs, we use AR/VR workloads that come prepackaged with the FARSI simulator~\cite{farsi}.
We use Pytorch2Timeloop~\cite{parashar2019timeloop} to convert the CNN models to a format that Timeloop accepts. Likewise, we use different CNN models available in Maestro to evaluate the optimal mapping.

\niparagraph{Agents.} 
For each agent, we took existing open-source implementations and modified the interfaces for our architectural design space problem. For ACO, we adopted Deepswarm~\cite{byla2019deepswarm}, which implements ant colony swarm intelligence. The skopt Python library~\cite{Scikit-opt} provides an implementation of GA Optimization. We use ACME~\cite{hoffman2020acme} research framework for RL. Our BO implementation was repurposed from the Scikit-opt~\cite{skopt} Python library. For  random walker implementation we used Numpy~\cite{numpy}.% to seed our agent's actions randomly. %The details about how we map various simulators to the \ours interface are presented in Table~\ref{tab:gymenv_summary}.

\section{Evaluation}
\label{sec:results}

%We first demonstrate how \ours is used to evaluate different ML algorithms for performing  DSE for three hardware architecture components: a DRAM memory controller, a DNN hardware accelerator, and an SoC. Next, we show how the performance of the ML algorithms varies under sample efficiency constraints. Lastly, we show that the dataset collected from the interface standardization for all ML algorithms can be beneficial to construct a proxy model of the simulator.
We demonstrate how \ours evaluates ML algorithms for architecture design space exploration in four hardware architecture components: a DRAM memory controller, a DNN hardware accelerator, an SoC, and DNN mapping problem. We then show how ML algorithm performance varies under different sample efficiency constraints. Finally, we demonstrate that the dataset collected from standardized interface from all ML algorithms can aid in constructing a high-fidelity proxy model of the simulator.

The insights from \ours are three-fold. First, all ML algorithms can equally find designs for a given target specification, but their performance is highly sensitive to hyperparameter selection. Therefore, finding optimal hyperparameters for each algorithm is akin to a lottery, and an exhaustive search reveals at least one set of parameters for each algorithm that achieves comparable performance (i.e., meets target design) to others.

Second, {normalizing comparisons using sample efficiency} is necessary, as it can be difficult to compare ML algorithms when each can achieve the best architectural solution given unlimited resources for hyperparameter optimization. Instead, by considering the constraints of sample efficiency, such as the number of samples that can effectively be collected from a given architectural simulator, a family of ML algorithms can be selected and compared effectively.

%Second, we need to use {sample efficiency as a comparison normalizer.} Even though under unlimited resources, it is possible to find a hyperparameter for each ML algorithm that achieves the best architectural solution, this also makes the comparison of ML algorithms hard. However, sample efficiency constraints (i.e., how many samples we can effectively collect from a given architectural simulator) can be used to select a family of ML algorithms effectively. %For instance, for architecture DSE problems where we are severely constrained by our ability to collect a large number of samples (e.g., less than 10000), even simple algorithms like random walker can be equally competitive to Bayesian optimization, evolutionary, or ACO. On the other hand, if we have unlimited access (e.g., greater than 250000) to architecture simulators (e.g., an analytical model-based or ML-based cost models), learning-based methods like reinforcement learning can be competitive compared to others.

Third, \ours provides standard interfaces used by all the ML algorithms. The information passed through this interface can create a standardized dataset. This feature ensures that each experiment results in a usable artifact,  irrespective of the type of ML algorithm. For instance, the diverse dataset collected from all ML agents for the same problem can be used to build a proxy model with better accuracy to overcome the sample efficiency problem of most slower architecture simulators.

\subsection{Hyperparameter Lottery and Domain Specific Operators}
\label{sec:hyper-lottery}
Using \ours, we demonstrate that across different optimization objectives and DSE problems, \textit{at least one} set of hyperparameters exists that results in the same performance as other ML algorithms.
A poorly selected (random selection) hyperparameter for the ML algorithm or its baseline can lead to a misleading conclusion that a particular family of  ML algorithms is better than another.
For instance for some algorithms finding the key set of hyperparameters is easy. Note that we might still need a large sweep to find the optimal values for the key hyperparameters. A case in point is using reinforcement learning or supervised learning algorithms with enough infrastructure and research momentum to identify the important set of hyperparameters. On the other hand, for the algorithms that are fallen out of the limelight, finding the right set of algorithms requires exhaustive search or even luck to make it competitive as its baseline.

\begin{comment}
\begin{figure}[t]
  \centering
  \includegraphics[width=1.0\columnwidth]{Arch-Gym/figs/FARSI_Table_Results.pdf}
  \caption{\rev{Redo: Parameters selected by each algorithm for the Edge Detection workload (which contains 8 subtasks) running on FARSI simulator. There are three high-level decisions in this design space to select: (1) allocation of resources (e.g., which PEs, memories, interconnects to instantiate in the SoC), (2) connection of resources (e.g., which PE/memory is connected to which bus), and (3) mapping of tasks to the PEs and memories (e.g., placement of tasks in time \& space onto resources). For sake of brevity, in our table we have only included decisions (1) and (3) of each algorithm. }.\vspace{-3em}}
  \label{fig:arch-gym-FARSI-results} 
  \vspace{1em}
\end{figure}
\end{comment}
To demonstrate the existence of the hyperparameter lottery, we compare the performance of each ML agent for the same architecture optimization problem. \ours provides a standardization interface which makes all the agent solve the problem using the same environment, target objective. This allows for a fair apples-to-apples comparison in terms of each agent's ability to solve the task. To benchmark the agent's performance, we compare it across the following axis: (1) How the target objective affect the ML agent's performance? (2) How the complexity of the architecture system affects the ML Agent's performance? To that end, we compare against three objectives namely, power, latency, and joint objective of minimizing latency and power. Likewise for comparing against increasing complexity of the architecture system, we vary from component-level, IP-level, and SoC level. For the component level, we want to design a custom DRAM memory controller for different workload traces. For the IP-level, we aim to design a custom neural network accelerator for different neural network architectures. Lastly, for the SoC level, we aim to design a custom DSSoC for different target workloads.

\niparagraph{Significant statistical variations.} \Fig{fig:dramsys-lottery} shows the comparison of different ML agents (named as ACO, BO, GA, Random Walker, RL) for the architecture DSE problem of finding optimal memory controller parameters for four different memory traces namely cloud-1, cloud-2, streaming access, and random memory access. We evaluate the performance of the ML agents for three different objectives: low power, low latency, and multi-objective optimization of latency and power. As shown in \Fig{fig:dramsys-lottery}, irrespective of the design objectives, we see a huge variance in the ML agents' performance depending upon the selected hyperparameter choice. In the worst case, there is up to 90\% statistical spread (measured as the interquartile range) across different workloads and target objectives.

\begin{figure}[t]
  \centering
  \includegraphics[width=1.0\columnwidth]{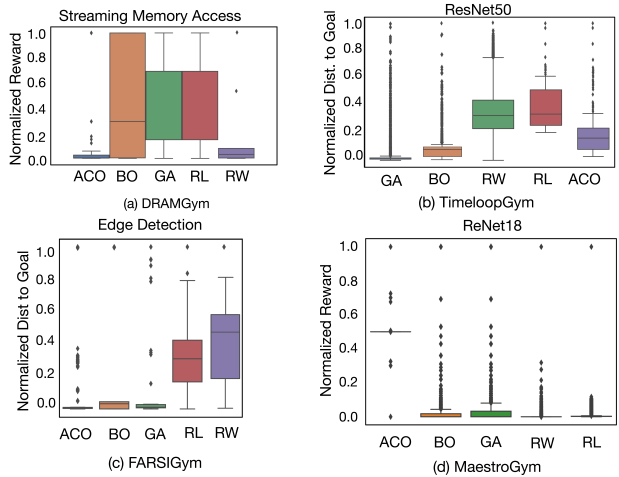}
  \caption{Hyperparameter lottery across different simulators and search algorithms for (a) DRAMGym, (b) TimeloopGym, (c) FARSIGym, and (d) MaestroGym. For TimeloopGym,  FARSIGym, and MaestroGym achieving lower distance or reward is better.}
  \label{fig:lottery-sims} 
\end{figure}

We observe the same trend across varying complexity of the architecture systems from component-level to SoC-level architecture exploration as shown in \Fig{fig:lottery-sims}. For example, we use the streaming access workload for the DRAM memory controller. For DNN hardware accelerator design, we search for an Eyeriss-like~\cite{chen2016eyeriss} hardware accelerator for the ResNet50 model. For SoC design, we use FARSI to evaluate different SoCs for edge detection workload. Likewise for DNN mapping, we find the best mapper for ResNet18 model.
Overall, on average, we perform more than 1.54 billion hardware simulations across 20 experimental setups (five for DRAMGym, TimeloopGym, FARSIGym, and MaestroGym respectively).
%\niparagraph{Implications:} The key takeaway is that across all agents, there is at least one hyperparameter configuration that can achieve a high reward/fitness to maximize the target objective. However, each agent also shows significant statistical variation. The variation is a byproduct of the exhaustive search of all available hyperparameters.% for the agent.

\niparagraph{Effectiveness of domain specific operators.}
\rev{We take GAMMA~\cite{gamma}, that uses MAESTRO~\cite{kwon2020maestro} for simulation, as an example.
GAMMA has introduced domain-specific operators, namely ``Aging'', ``Growth'', ``reordering''.
We compare the performance of GAMMA with four variants of genetic algorithms: 'GA-V1' (GA in GAMMA), 'GA+RO' (GA with reordering only), 'GA+AG' (GA with aging only), and 'GA+GR' (GA with growing).
In addition, we integrate MASTERO~\cite{kwon2020maestro} into \ours for comparisons with 'GA Arch-Gym', without domain-specific operators.
We perform an extensive hyperparameter sweep ($\sim$4000) experiments running for two days.}

\begin{figure}[t!]
  \centering
  \includegraphics[width=1.0\columnwidth]{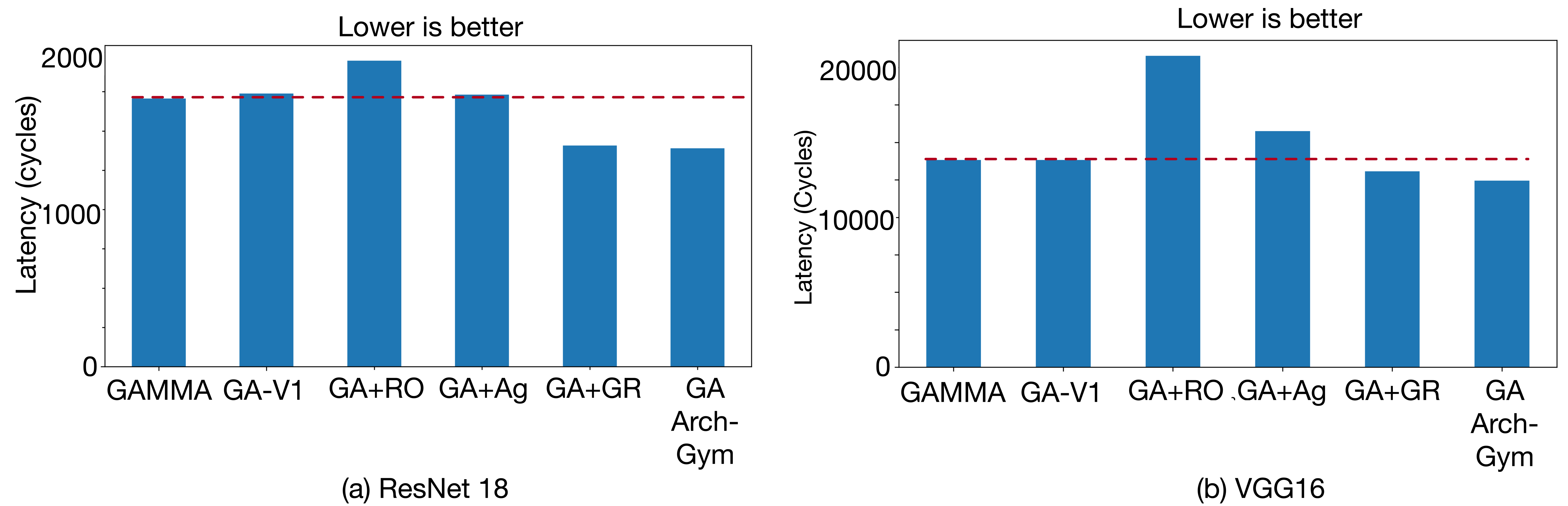}
  \caption{\rev{Comparision of latency of two ML models namely ResNet18 and VGG16 with GAMMA (with domain-specific operators) and vanilla genetic algorithm variants.  }}
  \label{fig:arch-gym-gamma} 
\end{figure}

\rev{\Fig{fig:arch-gym-gamma} summarizes the results for VGG16 and ResNet18.
The results illustrate that different variants of GA are equally effective in identifying the favorable design point.
%
% Our results show that all different variants achieve similar or better performance compared to GAMMA or any of its specialized operators (denoted by GA+RO, GA+Ag, and GA+GR).
Interestingly, GA in \ours, which does not have domain-specific operators, achieves better results than GAMMA.
These results further validate that when evaluating different search algorithms, it is critical to properly tune both the algorithm and its baselines, before making any conclusions about the efficiency of algorithms.}

%suggesting that the phenomenon is akin to the hardware lottery problem. 
\niparagraph{Implications.} The variation implies a number of implications. First, the choice of hyperparameter is critical not only for the ML agent of interest but also equally vital for other baselines when we compare the performance of different ML algorithms for architecture design space exploration. Second, after an exhaustive hyperparameter search of 21,600 experiments for these five ML agents, we found at least one hyperparameter configuration is equally competitive to other ML agents. Though, in all likelihood, we may have missed a lot of hyperparameter combinations which makes our subset of hyperparameters akin to winning a lottery. 

\niparagraph{A call to action.} The takeaway is that future ML-aided design requires us to report statistical distributions rather than report the state-of-the-art ML algorithm for a given architecture exploration problem. As we use these popular algorithms to tackle longstanding problems in architecture design space exploration, it is important to understand pitfalls. Otherwise, it is hard to operationalize the solutions in production and ensure industry adoption. Moreover, we must choose algorithms by considering domain challenges (e.g., scarcity in architecture datasets, the tradeoff between accuracy and speed with architecture simulators, etc.) rather than biasing towards any one algorithm approach since it is popular.

\begin{figure}[]
  \centering
  \includegraphics[width=1.0\columnwidth]{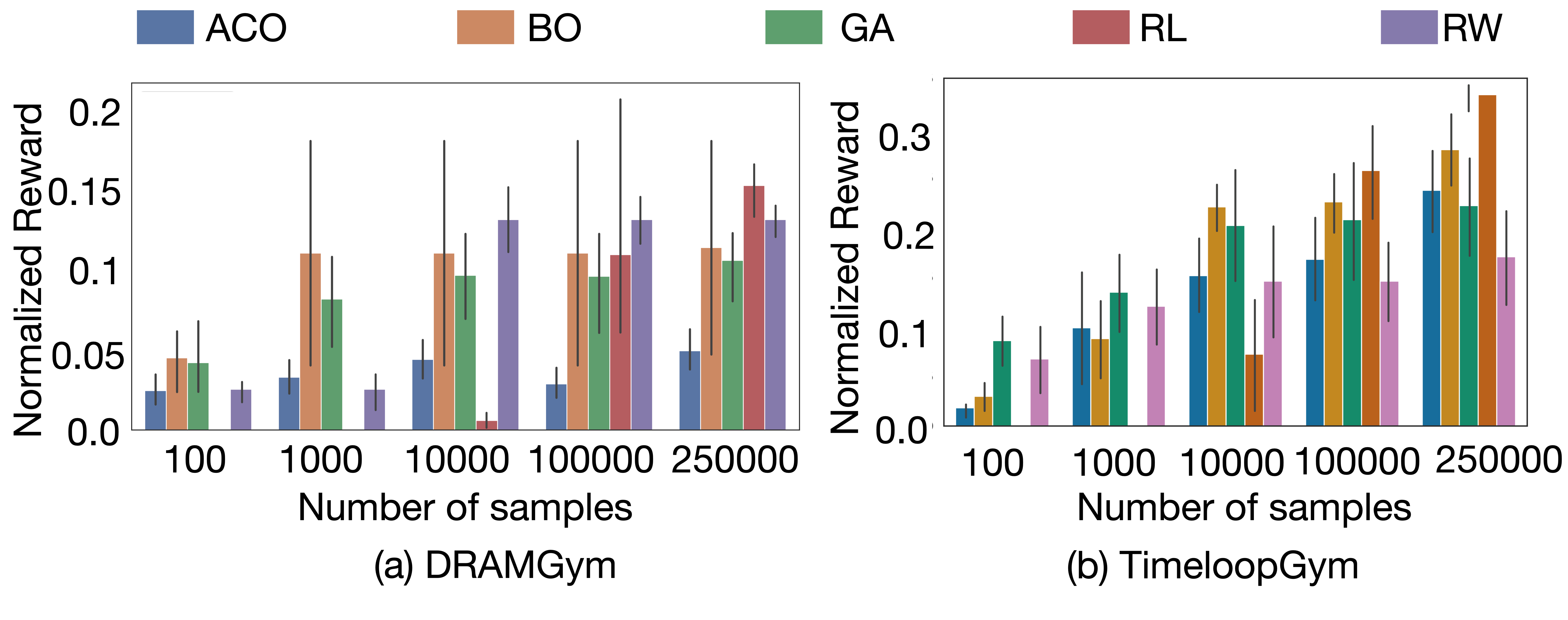}
  \caption{\rev{Mean normalized reward (target objective)} of ACO, BO, GA, RL, and RW for DRAM memory controller design (\rev{DRAMGym}) and \rev{ML accelerator design (TimeloopGym)} in a constrained setting, limiting the number of samples accessed by an
  algorithm from the simulator.\vspace{-3em}}
  \label{fig:arch-gym-tradeoffs} 
  \vspace{1em}
\end{figure}

\subsection{Trade-off Between ML Methods}
\label{sec:trade-offs}

Though in Section~\ref{sec:hyper-lottery}, we demonstrated that given a free run (unlimited number of resources to search), it is possible to find at least one solution that is equally good as the others.
However, as we consider the key challenges of the architecture domain, we observe certain trends that can be beneficial in selecting a particular ML algorithm for the architecture design space exploration. 
For instance, estimating the cost (e.g., power, latency, and area) for a given architecture parameter configuration can vary depending on the underlying architecture model, from analytical model-based (fast) to cycle-level (slow), and RTL simulation (extremely slow).
% The underlying architecture model can be analytical model-based, cycle-accurate simulators, transaction-level simulators, ML-based proxy cost models, or even worst-case RTL simulations or license-capped proprietary tools.
%

In such a scenario, how many samples we can effectively collect from the architecture model can be a useful constraint in determining different trade-offs in selecting a given ML algorithm for architecture design space exploration.

We use the number of samples from the architecture simulator as a normalization metric to compare the trade-offs between the search algorithms. We look at the number of samples we can query from simulator, in the range of 100, 1K, 100K, and 250K.

\Fig{fig:arch-gym-tradeoffs} compares different ML agents under sample efficiency constraints for the DRAM memory controller problem and DNN accelerator design. The \textit{x}-axis denotes the number of samples we can access from the simulator. The \textit{y}-axis denotes the mean normalized reward for each agent. We base our observation of the trade-offs on dividing the sample count into two regimes, namely the low sample count regime, and the high sample count regime.

\begin{figure}[t]
  \centering
  \includegraphics[width=1.0\columnwidth]{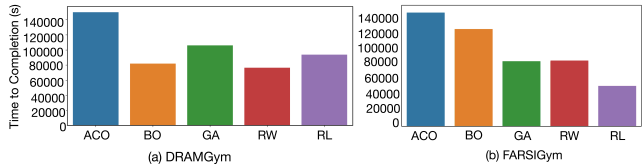}
  \caption{Comparison of ACO, BO, GA, RW, and RL in terms of time to completion for \rev{DRAMGym and FARSIGym}.}
  \label{fig:arch-gym-time-to-complete} 
  \vspace{-2em}
\end{figure}

\niparagraph{Low sample count.} In the low sample count regime ($\le$10000 samples), most ML agents perform decently well. Note that even simpler algorithms like Random walker (denoted as RW) are equally competitive to Bayesian optimization, genetic algorithm, and ant colony optimization. Finally, it is also worth noting that the performance of reinforcement learning is poor, as it is known that these algorithms are extremely sample inefficient.

\niparagraph{Higher sample count.} In the higher sample count regime ($\>$ 100000 samples), we observe that simpler algorithms (with exhaustive hyperparameter search) still remain competitive. However, we also see the performance of the reinforcement learning algorithm improve drastically compared to the lower sample count regime.

These results suggest that depending upon the speed of the architectural models collecting large samples can be prohibitively expensive; all popular algorithms except sample-inefficient algorithms like RL perform equally well. For an architecture model that trades speed vs. fidelity, where it is relatively easier to collect large data, we observe that emerging algorithms like reinforcement learning show increasingly better performance. However, it is important to note that other algorithms also remain equally competitive. 

\niparagraph{Time to Completion.} \Fig{fig:arch-gym-time-to-complete} illustrates the time to completion of various agents in DRAMGym and FARSIGym environment respectively. However, this comparison is not a fair assessment as it disregards the fact that some agents, such as ACO and GA, are multi-agent algorithms with different levels of optimization. ACO, for example, takes longer to complete as it relies on sequential evaluation, whereas GA benefits from parallel evaluation. Additionally, the run time of RL is comparable to that of RW and BO, despite the fact that RL is accelerated through GPU implementation.

For these reasons, we use sample efficiency, as a better comparison point across different ML agents. 
Additionally, sample efficiency, unlike other metrics, such as the time to completion and the number of hardware resources required depends upon the optimization effort to parallelize and finetune the ML agent implementation. 
In such cases, the number of samples from the simulator is a fair comparison because it directly considers the key domain challenges, such as architecture fidelity, speed, and licensing cost, to get useful data from the simulator.

\niparagraph{Implications.} The bottleneck in ML-aided architecture design space exploration is not the ML algorithm but the sheer slowness in the high-fidelity architecture cost model. On the one hand, slower architecture cost models make applying new emerging learning-based algorithms to architecture design space exploration harder. On the other hand, a faster architecture cost model allows sample inefficient learning-based algorithms (e.g., RL) to shine. 

\niparagraph{A call to action.}
Given an increased momentum toward novel learning-based algorithms like reinforcement learning and offline-RL~\cite{offline-rl}, we will likely see many different formulations and variants developed in the near future. Already, there are various learning frameworks~\cite{hoffman2020acme,espeholt2019seed,raffin2019stable,dhariwal2017openai,huang2021cleanrl,liang2018rllib,bauer2021reagent} built over popular ML research infrastructures like Tensorflow~\cite{abadi2016tensorflow} and Pytorch~\cite{paszke2019pytorch}. While novel learning algorithms formulations continue to use Atari-like games as a test bed, applying them to a real-world problem like architecture design exploration is challenging. Therefore, there is a need for open-source frameworks like \ours, that enables fair `apples-to-apples' comparison of the efficacy of rapidly evolving learning algorithms. Finally, a community-standard approach, akin to traditional cycle-level simulators like gem5 etc., allows dataset aggregation and high-quality dataset creation, which can help create more data-driven architecture cost models (e.g., proxy ML models).

\begin{table}[]
\centering
  \caption{\rev{Architectural parameters found by different search algorithms for finding a low-power DRAM memory controller (target goal: 1 Watt) design for a pointer chasing memory access pattern.}}
\resizebox{\linewidth}{!}{
\begin{tabular}{|c|ccccc|}
\hline
{\color{black} }                                     & \multicolumn{5}{c|}{{\color{black} \textbf{Parameter Values}}}                                                                                                                                                                                                               \\ \cline{2-6} 
\multirow{1}{*}{{\color{black} \textbf{Parameter}}} & \multicolumn{1}{c|}{{\color{black} \textbf{RL}}} & \multicolumn{1}{c|}{{\color{black} \textbf{RW}}} & \multicolumn{1}{c|}{{\color{black} \textbf{BO}}} & \multicolumn{1}{c|}{{\color{black} \textbf{GA}}}   & {\color{black} \textbf{ACO}} \\ \hline\hline
{\color{black} \textbf{Page Policy}}                 & \multicolumn{1}{c|}{{\color{black} Open Adaptive}}     & \multicolumn{1}{c|}{{\color{black} Open}}        & \multicolumn{1}{c|}{{\color{black} Open}}        & \multicolumn{1}{c|}{{\color{black} Open Adaptive}} & {\color{black} Open}         \\ \hline
{\color{black} \textbf{Scheduler}}                   & \multicolumn{1}{c|}{{\color{black} Fifo}}              & \multicolumn{1}{c|}{{\color{black} Fifo}}        & \multicolumn{1}{c|}{{\color{black} FrFcFs}}      & \multicolumn{1}{c|}{{\color{black} FrFcFs}}        & {\color{black} FrFcFs}       \\ \hline
{\color{black} \textbf{SchedulerBuffer}}             & \multicolumn{1}{c|}{{\color{black} Shared}}            & \multicolumn{1}{c|}{{\color{black} Shared}}      & \multicolumn{1}{c|}{{\color{black} Shared}}      & \multicolumn{1}{c|}{{\color{black} ReadWrite}}     & {\color{black} Bankwise}     \\ \hline
{\color{black} \textbf{Request Buffer Size}}         & \multicolumn{1}{c|}{{\color{black} 1}}                 & \multicolumn{1}{c|}{{\color{black} 4}}           & \multicolumn{1}{c|}{{\color{black} 4}}           & \multicolumn{1}{c|}{{\color{black} 1}}             & {\color{black} 4}            \\ \hline
{\color{black} \textbf{RespQueue}}                   & \multicolumn{1}{c|}{{\color{black} Reorder}}           & \multicolumn{1}{c|}{{\color{black} Fifo}}        & \multicolumn{1}{c|}{{\color{black} Reorder}}     & \multicolumn{1}{c|}{{\color{black} Reorder}}       & {\color{black} Fifo}         \\ \hline
{\color{black} \textbf{Refresh Max Postponed}}       & \multicolumn{1}{c|}{{\color{black} 4}}                 & \multicolumn{1}{c|}{{\color{black} 8}}           & \multicolumn{1}{c|}{{\color{black} 4}}           & \multicolumn{1}{c|}{{\color{black} 4}}             & {\color{black} 2}            \\ \hline
{\color{black} \textbf{Refresh Max Pulledin}}        & \multicolumn{1}{c|}{{\color{black} 8}}                 & \multicolumn{1}{c|}{{\color{black} 4}}           & \multicolumn{1}{c|}{{\color{black} 4}}           & \multicolumn{1}{c|}{{\color{black} 8}}             & {\color{black} 8}            \\ \hline
{\color{black} \textbf{Arbiter}}                     & \multicolumn{1}{c|}{{\color{black} Reorder}}           & \multicolumn{1}{c|}{{\color{black} Fifo}}        & \multicolumn{1}{c|}{{\color{black} Reorder}}     & \multicolumn{1}{c|}{{\color{black} Reorder}}       & {\color{black} Fifo}         \\ \hline
{\color{black} \textbf{Max Active Trans.}}           & \multicolumn{1}{c|}{{\color{black} 1}}                 & \multicolumn{1}{c|}{{\color{black} 1}}           & \multicolumn{1}{c|}{{\color{black} 1}}           & \multicolumn{1}{c|}{{\color{black} 1}}             & {\color{black} 1}            \\ \hline
\end{tabular}
\label{tab:arch-design-comparison}

}\
%\vspace{1em}
\end{table}

\subsection{Analysis of Designed Hardware}
Table~\ref{tab:arch-design-comparison} demonstrates the designed hardware for a DRAM memory controller across different agents.
We use a memory trace with random address access (e.g., pointer chasing).
The primay goal is to design a memory controller that achieves a power consumption of 1 Watt.
As shown in Table~\ref{fig:dramsys-lottery}, all the agents are able to find \textit{at least} one design that satisfies the target power consumption.
We observe that all agents keep the `Max Active Trans.' buffer size minimal with value of one.
Nonetheless, when the buffer sizes are different, for example `Request Buffer' or `Refresh Max Postponed', the agents reach to different `Page Policy', `Scheduler', and `SchedulerBuffer' in order to achieve the same power target of 1 Watt.

\section{Dataset Generation}
\label{sec:dataset}
Our results in Section~\ref{sec:trade-offs} show that a faster architectural model can unlock novel applications of learning-based algorithms such as reinforcement learning~\cite{kao2020confuciux} or data-driven offline reinforcement learning~\cite{yazdanbakhsh2021evaluation}. Indeed we notice the upward trend of using expensive learning-based algorithms~\cite{kao2020confuciux,yazdanbakhsh2021evaluation,drl-nocs,rl-transistor} for design space exploration that solely rely on analytical models~\cite{kwon2020maestro} or proxy ML-based cost models~\cite{yazdanbakhsh2021evaluation}, thus bypassing slow architectural simulators. 

However, like all statistical models, these ML-based proxy models are imperfect and suffer high prediction errors for out-of-distribution data~\cite{out-of-dist-error}. Thus, even though fast proxy models enable using learning-based algorithms (e.g., RL) for architecture design space exploration, the quality of designs generated from such architecture design space exploration needs more scrutiny. For instance, these proxy models continue to trade off accuracy for speed (see Section~\ref{sec:intro}). Hence, in this section, we answer the question of \textit{how to improve the accuracy of the proxy model while leveraging the gain in simulation speed}.

\begin{figure}[t!]
  \centering
  \includegraphics[width=1.0\columnwidth]{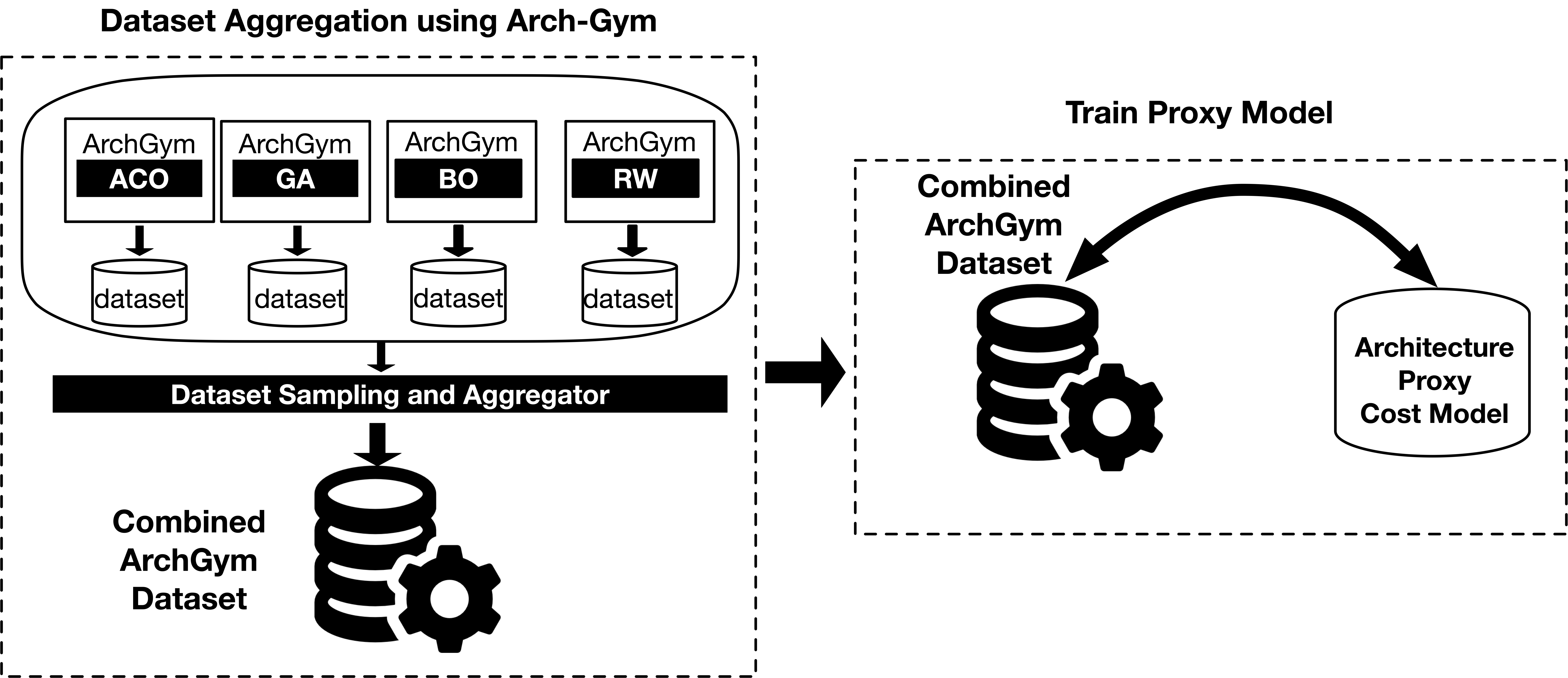}
  \caption{Dataset aggregation via \ours. Since all the ML agents use the same standardized interfaces, each experiment data can be leveraged to build a larger and diverse dataset.}
  \vspace{-2em}
  \label{fig:arch-gym-ds} 
\end{figure}

To that end, creating a unified interface using \ours for all ML agents also allows the creation of datasets that can be used to design better data-driven ML-based proxy architecture cost models to improve architecture simulator's speed. Using \ours datasets from DRAMGym explored across different ML agents, we construct an proxy cost model for predicting the latency, power, and energy.
Our results show that two things are important to improve the accuracy of the proxy models: \textit{Dataset size} and \textit{Dataset diversity}. \ours methodology seamlessly provides a means to improve dataset size and diversity.

\subsection{Dataset Construction}
\Fig{fig:arch-gym-ds} shows the dataset aggregation setup using \ours. The information exchange between the agent and the architecture environment is logged in a standardized dataset format~\cite{paper2021tensorflow,ramos2021rlds} for each hyperparameter exploration study. The standardized dataset can be seamlessly merged (for size) or sampled by an ML agent type (for diversity) to construct a high-quality, large, and diverse dataset.

 \begin{figure}
  \centering
  \includegraphics[width=0.8\columnwidth]{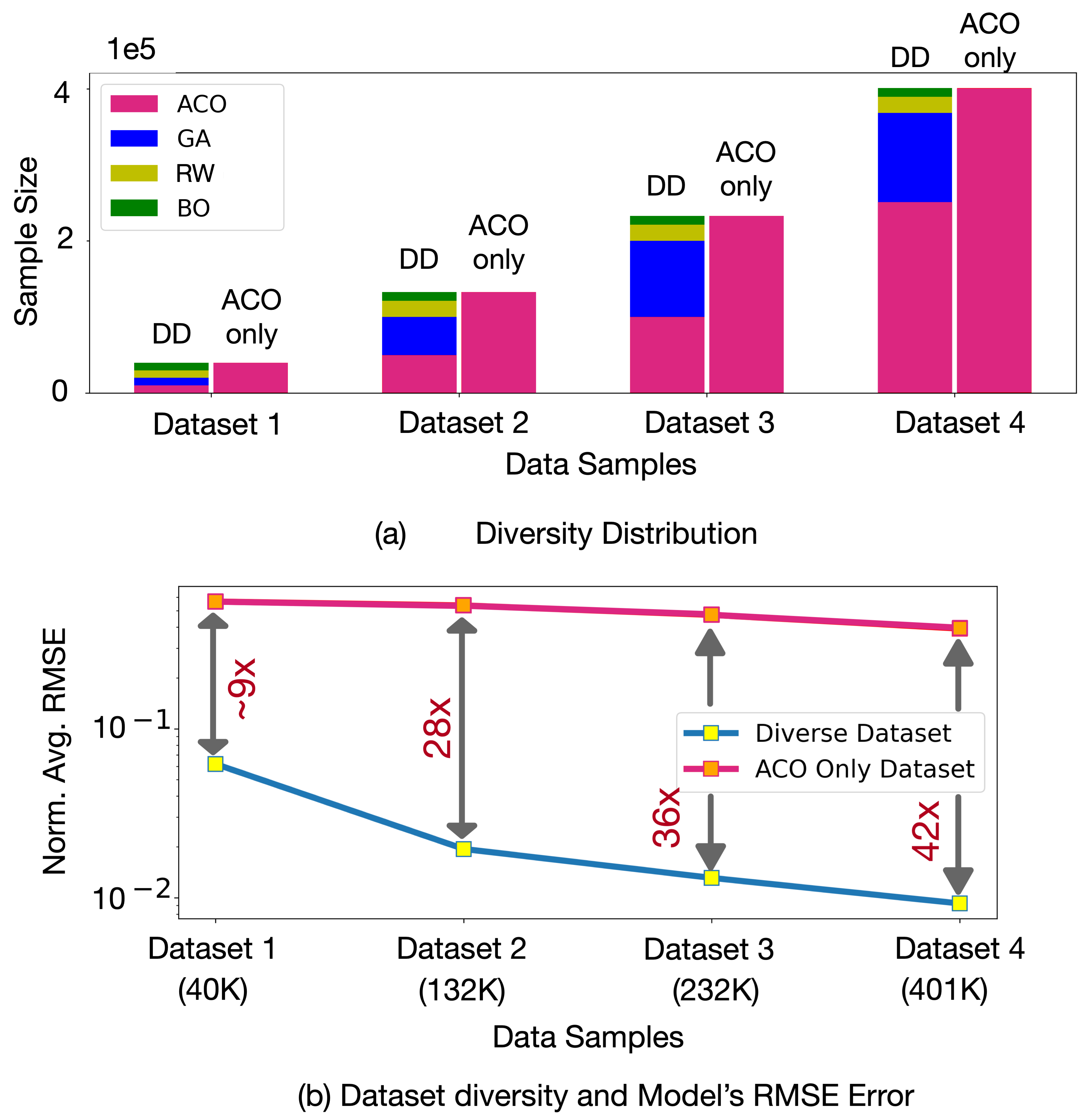}
  \caption{(a) Dataset characteristics and (b) its corresponding proxy model RMSE of the proxy model.}
  \label{fig:data_div_rmse} 
  \vspace{-1em}
\end{figure}

Using the setup shown in \Fig{fig:arch-gym-ds}, we construct four datasets, namely `Dataset 1', `Dataset 2', `Dataset 3', and `Dataset 4'. We categorize these four datasets into two groups: `Diverse dataset' (DD) and `ACO-only Dataset.' As the name suggests, the data is sourced from multiple agents in the Diverse dataset. This data comes from numerous hyperparameter explorations of ACO, GA, RW, and BO. \Fig{fig:data_div_rmse}-a shows the exact distribution in the composition of different datasets in Diverse dataset category. In the ACO-only case, the entire dataset is constructed only from the ACO's agent. Ideally, we can use the data from any other agent as well. Also, to construct the datasets of specific sizes (for example, Dataset 1 size $<$ Dataset 2 size), we use a random sampling utility in \texttt{pandas} to sample these two categories of datasets.

\subsection{High-Fidelity Proxy Model Training}
We use Random Forest~\cite{random_forest} model to predict the latency, power, and energy for the data collected from the DRAMGym environment. Each target (e.g., latency) is predicted by a separate random forest regression model. The features fed into the random forest model are the DRAMGym architecture parameters (see \Fig{fig:arch_dse_problem} for the list of parameters). We conducted a random hyperparameter search for each model across every dataset size to obtain models that achieves the lowest root mean square error (RMSE).

%\niparagraph{Problem With ML-Based Proxy Models.} To improve the performance of the proxy model intuitively, there is a need for a larger dataset. Moreover, these models also exhibit higher prediction errors for out-of-distribution data. Hence, to solve this problem, we need a way to collect large and diverse data.

\subsection{Implications}
\label{sec:takeaway}
Our results show that constructing a high-fidelity proxy cost model requires not just the quantity of the data but also the diversity of the dataset. To that end, we analyze the performance of the proxy cost model as we increase the dataset size without diversity as well as increasing the dataset size with diversity.

\niparagraph{Dataset size matters.} Our results demonstrate that the dataset size plays an important role in the accuracy of the proxy model as shown in \Fig{fig:data_div_rmse}-b (the trend line denoted for ACO Only Dataset). While this is intuitive, it is important to note that we still need to rely on slow architecture simulators to collect large datasets, which hinders large-scale data collection.

\ours tackles this problem better since it inherently facilitates the aggregation of large data from all the ML agents. Although one needs to run large sweeps due to the hyperparameter lottery, we believe each exploration experiment can be a useful artifact. Whether or not each run results in a better design is inconsequential since all these exploration data can be aggregated seamlessly due to its standardization provided by \ours. Prior work has shown that invalid designs and other sources of sub-optimal designs can be beneficial for architecture design space exploration~\cite{yazdanbakhsh2021evaluation}.

\niparagraph{Dataset diversity matters.} Our results demonstrate that dataset diversity also plays an important role in improving the accuracy of the proxy model as shown in \Fig{fig:data_div_rmse}-b (denoted by the trend line for Diverse Dataset). In fact, as the dataset size increases, the effect of sourcing data from diverse sources (in our case, different agents) is more pronounced. From aggregating data from DRAMGym, we observe that, on average, we can reduce the RMSE error by  42$\times$.

\begin{figure}[t]
  \centering
  \includegraphics[width=0.95\linewidth]{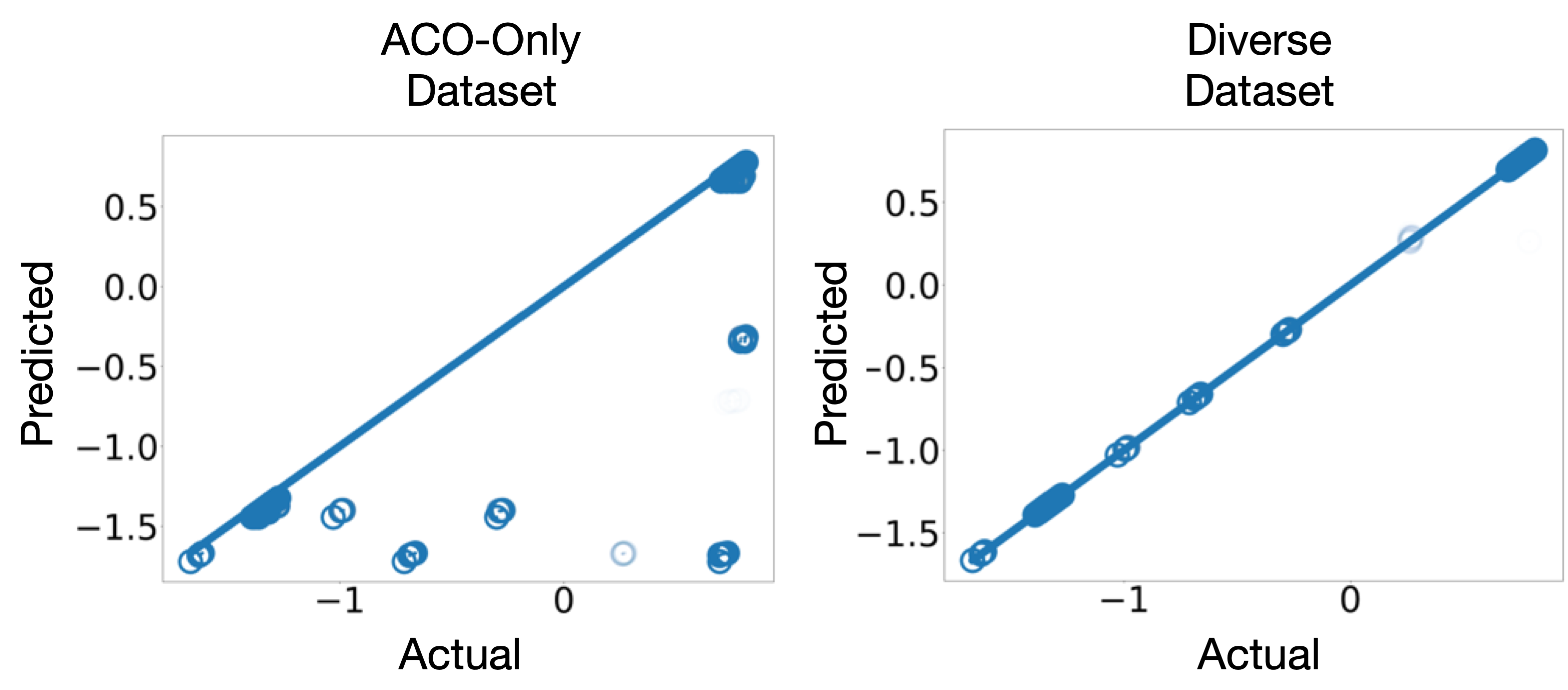}
  \caption{Comparison of actual vs predicted values for the power model between single source dataset and diverse dataset. For single source, we source all the data from one ML agent (ACO).}
  \label{fig:proxy_model_corr} 
\end{figure}

\Fig{fig:proxy_model_corr} visualizes the RMSE error, comparing the values predicted by the proxy models (for energy, latency, and power) and the actual ground truth values obtained from the DRAMSys simulator. 
We observe a consistent trend where the actual and predicted values are less correlated when we use a dataset from a single source across different proxy models. A diverse dataset helps improve coverage of the design space, thus resulting in a higher correlation. 

\begin{figure}
  \centering
  \includegraphics[width=\columnwidth]{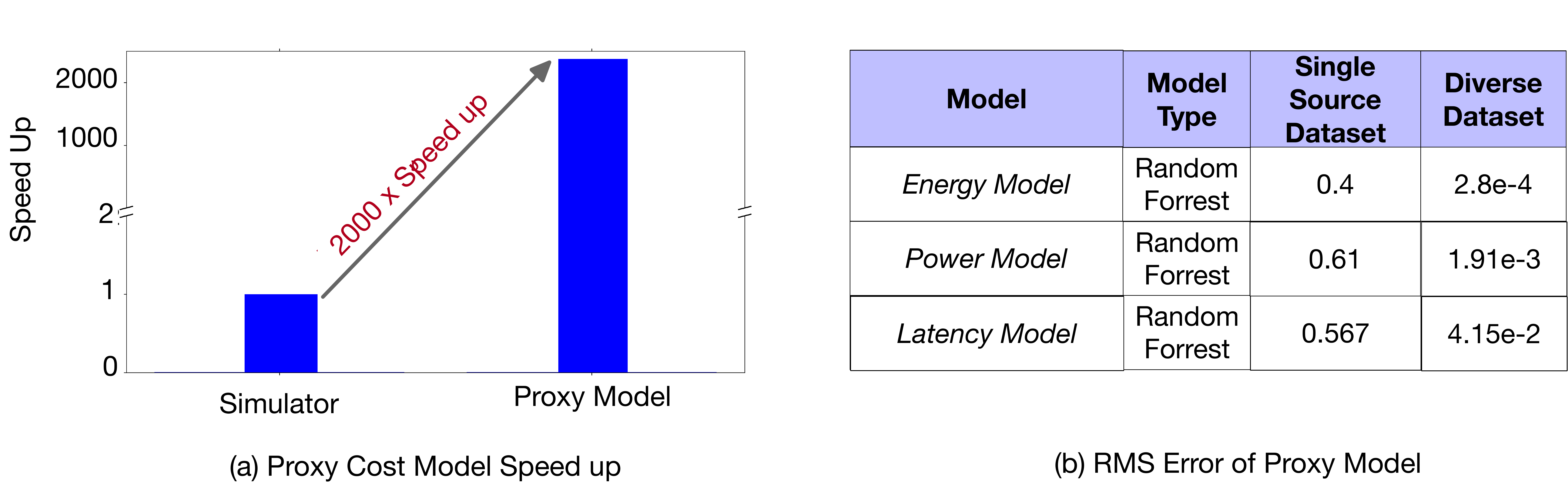}
  \caption{(a) Speed-up of the ML-based proxy cost model with cycle accurate simulator as the baseline. (b) The RMSE error for proxy cost model for latency, power, and energy.}
  \vspace{-1em}
  \label{fig:arch-gym-speed-up} 
\end{figure}

\niparagraph{Narrowing the gap between speed and accuracy.}
Another key insight of using \ours for dataset aggregation is that we can bridge the gap between the speed and accuracy of architecture cost models. 
Using the dataset aggregated from the DRAMGym environment using \ours, we show that the accuracy of the proxy model is comparable to the ground truth simulator while also resulting in 2000$\times$ speed up as shown in \Fig{fig:arch-gym-speed-up}. These high-accuracy yet speedy architecture cost models can be used to explore new and emerging learning-based algorithms such as reinforcement learning, offline-RL~\cite{offline-rl}, and multi-agent reinforcement learning, which were limited by the slowness in architectural simulators.   

While we demonstrate this using DRAMGym, we believe \ours natively provides the much-needed diversity in the dataset aggregation process for other slower architecture simulators. Furthermore, since all ML agents will have a different policy (see \Fig{fig:agent} in Section~\ref{sec:arch-gym-interface}), the way they explore the design space is also different. Thus, the exploration dataset aggregated through \ours across different ML agents (whether run by an individual researcher or community-wide adoption followed by aggregation) helps create a diverse dataset. These key features, in turn, improve the accuracy of the proxy model.

\section{Discussion on Extending \ours} 
\label{sec:archgym:extension}
\niparagraph{Integrating other proxy models.} We demonstrate how \ours can aid in creation of standardized and diverse datasets, which can be easily aggregated to balance the trade-off between accuracy and speed. By utilizing an accurate and high-speed proxy model, we can augment conventional slower architectural simulators while retaining their original interfaces. This enables us to leverage machine learning algorithms, including data-driven offline learning methods~\cite{kumar2021data} or offline reinforcement learning~\cite{offline-rl}, within \ours. Since \ours interfaces capture complex data, such as compiled IR or XLA graphs, we can train other deep learning models, such as GNNs~\cite{gnn-cost-model}, that achieve high accuracy. Regardless of the proxy model type, all models can be encapsulated using the same interface.

\niparagraph{Integrating other algorithms.}
\ours provides a unified interface not only for integrating hardware cost models, but also for integrating search algorithms or industry grade frameworks~\cite{tf-agents}.
Since each search algorithm (or agent) can be represented as a combination of a policy and hyperparameters (see Section~\ref{sec:arch-gym-interface}), any new search algorithm can be abstracted and integrated into \ours.
Additionally, unified search spaces, such as hardware-aware NAS~\cite{hw-nas}, require querying the hardware cost model (e.g., a simulator or real hardware) for state and receiving feedback based on optimization objectives. These problems can also be mapped into \ours (see Figure~\ref{fig:arch-gym-design}).
\section{Conclusion}
\label{sec:conclusion}
\ours is an open source gymnasium for ML-aided architecture design space exploration. Using our framework, we show the existence of the hyperparameter lottery. Moreover, \ours provides a standardized interface that can be extended and allows for fair comparison between different ML algorithms for a given architecture exploration problem. The standard interfaces for all ML algorithms further enable the creation of diverse datasets that can be used to explore novel, data-driven offline learning algorithms and build fast architectural cost models with high-fidelity.

\section*{Acknowledgments}

The authors would like to thank anonymous reviewers for their feedback on improving the manuscript's quality. 
Specifically, we would like to thank Sheng-Chun Kao for his help in understanding GAMMA codebase. 
We would like to thank Prof. Radhika Nagpal for helping us understand Ant-conlony optimization and Genetic algorithms.
We would also like to thank Natasha Jaques, Shayegan Omidshafiei, Izzeddin Gur from Google Research for helpful discussions on reinforcement learning.
We extend our gratitude towards Cliff Young, James Laudon, and extended Google Research, Brain Team for
their feedback and comments. We also thank Douglas Eck and Hardik Sharma for feedback on the early draft of this work.
This work is supported by the Office of the Director of National Intelligence (ODNI), Intelligence Advanced Research Projects Activity (IARPA), via 2022-21102100013. The views and conclusions contained herein are those of the authors and should not be interpreted as necessarily representing the official policies, either expressed or implied, of ODNI, IARPA, or the U.S. Government. The U.S. Government is authorized to reproduce and distribute reprints for governmental purposes notwithstanding any copyright annotation therein.



%\appendix
%\input{Arch-Gym/tex/artifact}

\bibliographystyle{ACM-Reference-Format}
%\balance
\bibliography{Arch-Gym/refs}

\end{document}